\documentclass[showpacs, pra, aps, twocolumn,10pt]{revtex4}
\usepackage{mathrsfs}
\usepackage{array}
\usepackage{amsmath}
\usepackage{graphicx}
\usepackage{amstext}
\usepackage{bm}
\usepackage{amsfonts}

\begin{document}

\title{Quantum-feedback-controlled macroscopic quantum nonlocality in cavity optomechanics}
\author{Yaqin Luo}
\affiliation{Department of Physics, Huazhong Normal University, Wuhan 430079, China}
\author{Huatang Tan}
\email{tht@mail.ccnu.edu.cn}
\affiliation{Department of Physics, Huazhong Normal University, Wuhan 430079, China}
\begin{abstract}
In this paper, we propose a continuous measurement and feedback scheme to achieve strong Einstein-Podolsky-Rosen (EPR) steering and Bell nonlocality of two macroscopic mechanical oscillators in cavity optomechanics. Our system consists of two optomechanical cavities in which two cavity fields are coupled to each other via a nondegenerate parametric downconversion. The two cavity output fields are subject to continuous Bell-like homodyne detection and the detection currents are fed back to drive the cavity fields. It is found that when the feedback is absent, the two mechanical oscillators can only be prepared in steady weakly entangled states which however do not display EPR steering and Bell nonlocality, due to the so-called 3 dB limit. But when the feedback is present, it is found that the mechanical entanglement is considerably enhanced such that strong mechanical steering and Bell nonlocality can be obtained in the steady-state regime. We analytically reveal that this is because the feedback drives the mechanical oscillators into a steady approximate two-mode squeezed vacuum state, with arbitrary squeezing in principle. It is shown that the feedback can also obviously improve the purity of the nonclassical mechanical states. The dependences of the mechanical quantum nonlocality on the feedback strength and thermal fluctuations are studied, and it is found that Bell nonlocality is much more vulnerable to thermal noise than EPR steerable nonlocality.
\end{abstract}

\maketitle

\section{Introduction}
In the year of 1935, Einstein, Podolsky, and Rosen (EPR) put forward the famous EPR paradox to demonstrate the incompatibility between the concepts of local causality
and the completeness of quantum mechanics. The EPR paradox involves the situation where two distant observers Alice and Bob
share an entangled particles and one observer, say, Alice, is able to control the
states of Bob's particle by performing local measurements on her particle \cite{EPR}. In his response to the EPR paradox, Schr\"{o}dinger termed this kind of ability to nonlocally control states of remote particles as steering \cite{Str}.  Based on intuitively reasonable notions of reality and locality, Bell in 1964 formulated the well-known Bell's inequality which provides a resolution of the EPR paradox in the sense that the violation of the inequality implies the failure of local realism but the existence of nonlocal quantum correlations (Bell nonlocality)\cite{bell1}.

EPR steering also characterizes another kind of quantum nonlocal effect. Very recently,  Wiseman, Jones, and Doherty have revisited the concept of quantum steering by regarding it as verifiable entanglement
distribution by an untrusted party \cite{wiseman01, wiseman02}. It was showed that EPR steering is intermediate between entanglement (inseparability) \cite{Horodecki}
and Bell nonlocality, the latter two defined respectively as entanglement distribution between trust parties and among distrust parties. Therefore,
the states display Bell nonlocality are a subset of steerable states which are, in turn, a subset of inseparable states. Distinct from entanglement and Bell nonlocality, steering is intrinsically asymmetric with respect to the two observers and it is thus directional.
Experimental verification of EPR steering and Bell nonlocality has already been achieved in a variety of atomic and photonic systems \cite{bell,schnabel, wollmann, cfli2, cfli3, ou, bowen, wiseman04, cfli, str1, str2, pryde, guerreiro, qyhe03, Walborn}. Besides being of fundamental interest, EPR steering and Bell nonlocality can also be considered as potential resource for quantum information processing, such as device-independent quantum cryptography \cite{bar, qucry1, qucry2} and secure quantum teleportation \cite{qyhe}. By utilizing steering, one can also achieve desirable quantum states by local measurements \cite{shen,tan}.

On the other hand, in past decade cavity optomechanics has witnessed rapid development \cite{opm}. Recent experiments have already achieved squeezed states of light and mechanical oscillators \cite{squ1, squ2}, light-mechanical entangled states \cite{omen}, nonclassical correlations between single photons and phonons from a mechanical oscillator \cite{markus1}, and even entangled states of two optomechanical oscillators \cite{markus2, me}. Cavity optomechanical system has now become as a prime indicate for studying various quantum phenomena on the macroscopic scale \cite{lv,li}. The generation of EPR steering of two macroscopic mechanical oscillators by pulsed cavity optomechanics was investigated \cite{qy}. Very recently, optomechanical Bell nonlocality has been experimentally demonstrated \cite{bell3}.

In this paper, we consider the realization of steady-state EPR steering and Bell nonlocality of two macroscopic mechanical oscillators by utilizing quantum feedback. It should be  noted that feedback has been frequently employed in cavity optomechanical systems for achieving e.g. cooling, normal-mode splitting, and even mechanical squeezing \cite{sch, chen, rossi, vin}. Our system consists of two optomechanical cavities in which the two cavity fields interact to each other by parametric downconversion and are also coupled respectively to a mechanical oscillator. We further consider that the two cavity output fields are subject to Bell-like homodyne detection and the detection currents are fed back to drive the cavity fields and alter the dynamics of the whole system. We show when the feedback is absent, the two mechanical oscillators can be prepared in steady entangled states, via the downconversion interaction, which, however, do not exhibit quantum steering and Bell nonlocality. But when the feedback is present, we find that under some conditions the two mechanical oscillators can be driven into an approximate two-mode squeezed vacuum steady state which displays strong quantum steerable correlations and Bell nonlocality. This feature make the present scheme obviously different from the existing quantum-reservoir-engineering scheme in Ref.\cite{jie}, in which only mixed mechanical entangled state which just displays some kind of Bell-nonlocality in short-time regime can be achieved.  In addition, the effect of thermal noise is discussed and it is shown that the Bell nonlocality is much more vulnerable to thermal fluctuations than the steering. Apart from the potential application in fundamental tests, e.g., quantum-to-classical transition \cite{zurek} and quantum mechanics in the regime of macroscopic objects, the present scheme may also be used for quantum information processing, like quantum communications based on micromechanical resonators \cite{markus2, me}.

\begin{figure}[t]
\centerline{\scalebox{0.31}{\includegraphics{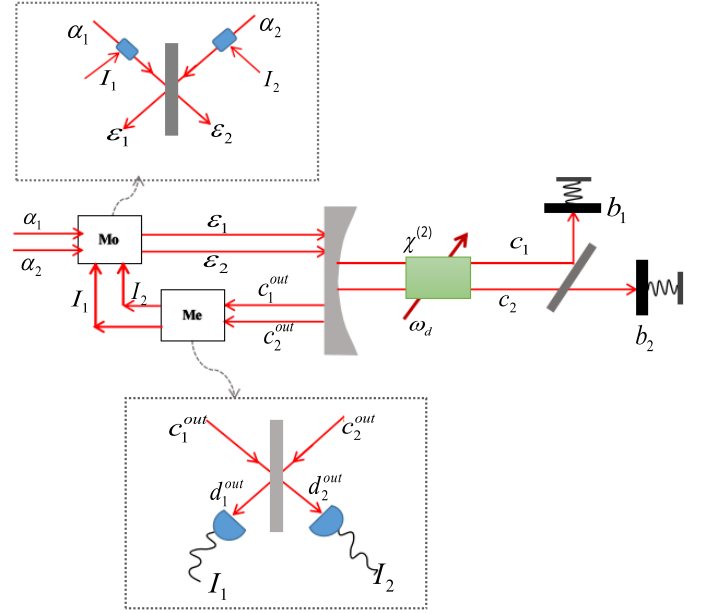}}}
 \caption{The schematic plot of the cavity optomechanical system and feedback loop. The two cavity fields $c_j~(j=1,2)$ (e.g., nondegenerate polarization) interacts with each other via an intracavity nondegenerate parametric downconversion (driven by a traveling laser field of frequency $\omega_d$) and are respectively coupled to a mechanical oscillator $b_j$. The cavity output fields $c_j^{\rm out}$ are mixed at a balanced beam splitter and the two outputs $d_j^{\rm out}$ from the beam splitter are subject to homodyne detection (denoted by the symbol ``Me"). The detection currents $I_j$ are fed back to drive the cavity fields via modulating the driving
 laser fields $\alpha_j$ (denoted by the symbol ``Mo").}
 \label{fig1}
\end{figure}

The remainder of this paper is organized as follows. In Section II, the model is introduced and the working equations are derived. In Section III, the criteria of two-mode Gaussian entanglement, steering, and Bell nonlocality are reviewed in brief. In Section IV, we present the results and discussion of the effects of quantum feedback on the entanglement, steering and Bell nonlocality of two optomechanical oscillators. In the last Section V, we give the main summary.

\section{System and equations}
\subsection{Optomechanical system}
As schematically shown in Fig.\ref{fig1}, we consider an optomechanical system in which two laser-driven optomechanical cavities are dispersively coupled to a mechanical oscillator, respectively. The two cavity fields interact with each other via nondegenerate parametric downconversion (NPD) with second-order optical nonlinearity. We note that the quantum properties, e.g., mechanical squeezing and cooling, of an optomechanical cavity containing a driven degenerate parametric downconvertor have been studied \cite{huang, ben}.
We assume that the cavity fields have the resonant frequencies $\omega_{cj}~(j=1,2)$ and the mechanical oscillators possess the  frequencies of $\omega_{mj}$. The cavity fields are respectively driven by lasers with frequencies $\omega_{lj}$ and amplitudes $\mathcal E_j~(j=1,2)$.
The Hamiltonian of the whole optmechanical system can be written as ($\hbar=1$)
\begin{align}
\hat H=&\sum_{j=1}^{2}\Big[\omega_{c_j}\hat{C}_j^\dag\hat{C}_j+\omega_{m_j}\hat{B}_j^\dag\hat{B}_j-g_{v_j }\hat{C}_j^\dag\hat{C}_j\big(\hat{B}_j^\dag +\hat{B}_j\big)
\nonumber\\
  &+i\mathcal{E}_j\big(\hat{C}_j^\dag\ e^{-i\omega_{lj}t}- \hat{C}_j\ e^{i\omega_{lj}t}\big)\Big]\nonumber\\
  &+2ig_p\big(\hat{C}_1^\dag \hat{C}_2^\dag\ e^{-i\omega_dt}-\hat{C}_1 \hat{C}_2 e^{i\omega_dt}\big),
  \end{align}
where the annihilation operators $\hat{C}_j$ and $\hat{B}_j$ are respectively describe the $j$th cavity field and mechanical oscillator, $g_{v_j}$ characterize single-photon optomechanical coupling strengthes, and $g_p$ represents the strength of the NPD pumped at frequency $\omega_d$. When $\omega_d=\sum_j\omega_{lj}+\omega_{mj}$, in the rotation frame with respect to the driving frequency $\omega_{l}$, the above Hamiltonian reduces to
\begin{align}
\hat H=&\sum_{j=1}^{2}\Big[\delta_{0j}\hat{C}_j^\dag\hat{C}_j+\omega_{mj}\hat{B}_j^\dag\hat{B}_j-g_{vj}\hat{C}_j^\dag\hat{C}_j\big(\hat{B}_j^\dag +\hat{B}_j\big)
\nonumber\\
  &+i\mathcal{E}_j\big(\hat{C}_j^\dag- \hat{C}_j\big)\Big]\nonumber\\
  &+2ig_p\big[\hat{C}_1^\dag \hat{C}_2^\dag\ e^{-i(\omega_{m1}+\omega_{m2})t}-\hat{C}_1 \hat{C}_2 e^{i(\omega_{m1}+\omega_{m2})t}\big],
  \label{ham2}
  \end{align}
where the detuning $\delta_{0j}=\omega_{cj}-\omega_{lj}$. For strong external driving fields, the above Hamiltonian (\ref{ham2}) can be linearized around the classical amplitudes $\langle  C_j\rangle$ and $\langle  B_j\rangle$ of the cavity and mechanical modes by expanding the operators $\hat{C}_j=\langle  \hat C_j\rangle+ \hat{c}_j$ and $\hat B_j=\langle \hat B_j\rangle+\hat b_j$, with $|\langle  C_j\rangle|^2\gg \langle\hat{c}_j^\dag\hat{c}_j\rangle$ and $|\langle  B_j\rangle|^2\gg \langle\hat{b}_j^\dag\hat{b}_j\rangle$. Here, the operators $\hat{c}_j$  and $\hat{b}_j$ represent quantum fluctuations of the cavity fields and the mechanical oscillators. According to Eq.(\ref{ham2}) and taking into account the cavity dissipation and mechanical damping, the classical amplitudes $\langle \hat C_j\rangle$ and $\langle  \hat B_j\rangle$ are satisfied by the equations $\frac{d}{dt}\langle\hat C_j\rangle\approx -(\kappa+i\Delta_j)\langle\hat C_j\rangle+\mathcal{E}_j$ and $\frac{d}{dt}\langle\hat B_j\rangle=-\big[\frac{\gamma_{m}}{2}+i\omega_{mj}\big]\hat B_j+ig_{vj}|\langle\hat{C}_j\rangle|^2$, where the detuning $\Delta_j=\delta_{0j}-2g_{vj} \text {Re}[\langle \hat B_j\rangle]$, and $\kappa$ and $\gamma_{m}$ are respectively the rates of cavity losses and the mechanical damping. The steady-state values $\langle\hat C_j\rangle_{ss}=\frac{\mathcal E_j}{\kappa+i\Delta_j}$ and $\langle\hat B_j\rangle_{ss}=\frac{2g_{vj}|\langle C_j\rangle_{ss}|^2}{2\omega_{mj}-i\gamma_m}$. Then, the linearized Hamiltonian can be obtained as
\begin{align}
  \hat H_{\rm lin}=&\sum_{j=1}^2\Big[\Delta_j \hat{c}_j^\dag\Hat{c}_j+\omega_{mj}\Hat{b}_j^\dag\Hat{b}_j+ig_{oj}( \Hat{c}_j-\Hat{c}_j^\dag)(\Hat{b}_j+\Hat{b}_j^\dag)\Big]\nonumber\\
  &+2ig_p\big[\hat{c}_1^\dag  \hat{c}_2^\dag\ e^{-i(\omega_{m1}+\omega_{m2})t}- \hat{c}_1 \hat{c}_2 e^{i(\omega_{m1}+\omega_{m2})t}\big],
  \end{align}
where the linear optomechanical coupling strength $g_{oj}=g_{v_j} |\langle\hat C_j\rangle_{ss}|$. If choosing the detuning $\Delta_j=\omega_{mj}$, i.e.,  the two cavity modes are driven resonantly on the red-detuned sidebands, one can find that the above Hamiltonian reduces to
\begin{align}
  \hat H_{\rm lin}=&\sum_{j=1}^2\Big[(ig_{oj} \hat c_j\hat b_j^\dag-ig_{oj} \hat c_j^\dag\hat b_j+ig_{oj} \hat c_j\hat b_j e^{-2i\omega_{mj}t}\nonumber\\
  &-ig_{oj} \hat c_j^\dag\hat b_j^\dag e^{2i\omega_{mj}t}\Big]+2ig_p(\hat{c}_1^\dag  \hat{c}_2^\dag- \hat{c}_1  \hat{c}_2).
  \label{effem2}
  \end{align}
When the mechanical frequency $\{\omega_{mj}\gg g_{oj}, \kappa, \gamma_m\}$, the fast-oscillating exponential terms in the above Hamiltonian can be neglected under the rotation-wave approximation (RWA), and then the Hamiltonian (\ref{effem2}) becomes into
\begin{align}
\hat H_{\rm lin}=&\sum_{j=1}^2 ig_{oj}(\hat c_j\hat b_j^\dag-\hat c_j^\dag\hat b_j)+2ig_p(\hat{c}_1^\dag  \hat{c}_2^\dag- \hat{c}_1  \hat{c}_2).
\label{linh}
\end{align}
Therefore, our scheme is also suitable for two different mechanical oscillators with different frequencies. It can be seen from the above equation that the steady-state entanglement between the two cavity modes can be built up via the NPD process and at the same time the intracavity-field entanglement is transferred to the two mechanical oscillators via the light-mechanical linear mixing described by the first term. However, it is well known that the inevitable cavity dissipation considerably limits the steady-state nonclassical properties of the two cavity fields, e.g., the maximally achievable two-mode squeezing is about fifty percent with respect to vacuum fluctuations (the so-called 3 dB limit), which in turn constraints the intracavity field entanglement and also the entanglement between the two mechanical oscillators. In addition, it has already been shown by one of us in Ref.\cite{tan} that the steady-state steering via the NPD with equal dissipation rates of two fields is unattainable. Therefore, in the following we consider using quantum feedback to exceed this limit and enhancing the mechanical entanglement to achieve mechanical steering and Bell nonlocality in the steady-state regime.


\subsection{Adding quantum feedback}
 As depicted in Fig.\ref{fig1}, the feedback consists of continuous monitoring of the cavity output fields and sending the detection results back to drive the cavity fields to modify the dynamics of the whole optomechanical system. Here, we consider a Bell-like detection of the two cavity output fields $\hat c_j^{\rm out}(t)\equiv\sqrt{2\kappa}\hat c_j(t)+\hat c_j^{\rm in}(t)$. The two cavity output fields $\hat c_1^{\rm out}(t)$ and $\hat c_2^{\rm out}(t)$ are combined at a balanced beam splitter and the two beam-splitter output fields,
 \begin{subequations}
 \begin{align}
 \hat d_1=\frac{1}{\sqrt{2}}(\hat c_1^{\rm out}(t)+\hat c_2^{\rm out}(t)),\\
\hat d_2=\frac{1}{\sqrt{2}}(\hat c_1^{\rm out}(t)-\hat c_2^{\rm out}(t)),
 \end{align}
 \end{subequations}
are under homodyne detection. Here we consider that the phase quadrature $\hat Y_{d_1}$ and amplitude quadrature $\hat X_{d_2}$ are under homodyne detection, where the quadratures
 \begin{subequations}
 \begin{align}
 \hat X_{\mathcal O}&=\frac{1}{\sqrt{2}}(\hat {\mathcal O}+\hat {\mathcal O}^\dag),\\
 \hat Y_{\mathcal O}&=-\frac{i}{\sqrt{2}}(\hat {\mathcal O}-\hat {\mathcal O}^\dag),
 \end{align}
 \end{subequations}
 for some bosonic operator $\hat {\mathcal O}$ (similarly hereinafter). In this way, the corresponding detection results (currents) can be described by the operators $\hat J_1$ and $\hat J_2$, which are given by \cite{wiseman}
\begin{subequations}
\begin{align}
\hat J_1(t)&=\frac{1}{\sqrt{2}}(\hat Y_{c_1}+\hat Y_{c_2})+\frac{\hat W_1(t)}{\sqrt{2\kappa\eta_f}},\\
\hat J_2(t)&=\frac{1}{\sqrt{2}}(\hat X_{c_1}-\hat X_{c_2})+\frac{\hat W_2(t)}{\sqrt{2\kappa\eta_f}}.
\end{align}
\end{subequations}
The parameter $\eta_f$ accounts for the detection efficiency and the noises $\hat W_j(t)$ due to the detection satisfy the following nonzero correlations
 \begin{align}
\langle W_j(t) W_{j'}(t')\rangle=\frac{1}{2}\delta_{jj'} \delta (t-t'),
\label{nc1}
 \end{align}
\begin{subequations}
\begin{align}
&\langle \hat W_1(t)\hat X_{c_1}^{\rm in}(t')\rangle=\langle \hat W_1(t)\hat X_{c_2}^{\rm in}(t')\rangle\nonumber\\
&~~~~~~~~~~~~~~~~~~~=-\langle \hat W_2(t)\hat Y_{c_1}^{\rm in}(t')\rangle=\langle \hat W_2(t)\hat Y_{c_2}^{\rm in}(t')\rangle\nonumber\\
&~~~~~~~~~~~~~~~~~~~=-\frac{i\sqrt{\eta_f}}{2\sqrt{2}}\delta(t-t'),\\
\text{and}\nonumber\\
&\langle \hat W_2(t)\hat X_{c_1}^{\rm in}(t')\rangle=-\langle \hat W_2(t)\hat X_{c_2}^{\rm in}(t')\rangle\nonumber\\
&~~~~~~~~~~~~~~~~~~~=-\langle \hat W_1(t)\hat Y_{c_1}^{in}(t')\rangle=\langle \hat W_1(t)\hat Y_{c_2}^{\rm in}(t')\rangle\nonumber\\
&~~~~~~~~~~~~~~~~~~~=\frac{\sqrt{\eta_f}}{2\sqrt{2}}\delta(t-t').
\end{align}
\label{nc2}
\end{subequations}
The above equations (\ref{nc2}) show that when the detection efficiency $\eta_f=0$, the detection noises $\hat W_j(t)$ and the cavity input noises $\hat X_j(t)$ and $\hat Y_j(t)$ become independent, which means the absence of the feedback. For the efficiency $\eta_f=1$, which means that all the light lost by the cavity is detected and employed in the feedback loop, the feedback is optimal and can in principle be realized via reflection with one-sided cavities \cite{zip}.

The currents are then fed to drive the cavity fields in the way described by the following Hamiltonian
\begin{subequations}
  \begin{align}
 \hat  H_{f_1}&=\frac{\lambda_{f_1}}{\sqrt{2}} \hat J_1(t-\tau_1)(\hat X_{c_1}+\hat X_{c_2}),\\
 \hat  H_{f_2}&=-\frac{\lambda_{f_2}}{\sqrt{2}} \hat J_2(t-\tau_2)(\hat Y_{c_1}-\hat Y_{c_2}),
  \end{align}
  \label{11}
  \end{subequations}
where $\lambda_{f_j} $ represents the corresponding feedback gains and $\tau$ the feedback loop delay time. When considering symmetric parameters, the feedback strengthes $\lambda_{f_j} $ is also symmetric with respect to the two optomechanical subsystems and we thus set $\lambda_{f_j}=\lambda_f$. In addition, when the delay time is much shorter than the characteristic time of the system, the time delay can be negligible, i.e., $\tau\rightarrow0$, which means that Markovian feedback is taken into account. Very recently, by the feeding the homodyne current back to the cavity field, experiments have realized the enhanced cooling and normal-mode splitting in optomechanical systems \cite{rossi}.

Including the feedback Hamiltonian in Eq.(\ref{11}) and taking into account of the cavity dissipation and mechanical damping, the equations of motion  for the system's operators $\hat b_j$ and $\hat c_j$ can be derived as
\begin{subequations}
\begin{align}
  \frac{d}{dt}\hat{b}_j=&-\frac{\gamma_m}{2} \hat{b}_j+g_{oj}(\hat{c}_j-\hat{c}_j^\dag e^{2 i\omega_mt})+\sqrt{\gamma_m}\hat{b}_j^{\rm in}(t),\\
  \frac{d}{dt}\hat c_1=&-\big(\kappa+\frac{\lambda}{2}\big)\hat{c}_1+(2g_{p}+\frac{\lambda}{2})\hat{c}_2^\dag-g_{ o}(\hat{b}_1+\hat{b}_1^\dag e^{2i\omega_mt})\nonumber\\
  &+\sqrt{2\kappa}\hat{c}_1^{\rm in}(t)-\frac{\lambda}{2\sqrt{2\kappa\eta_f}}(i\hat W_1(t)+\hat W_2(t)),\\
  \frac{d}{dt}\hat c_2=&-(\kappa+\frac{\lambda}{2})\hat{c}_2+(2g_{p}+\frac{\lambda}{2})\hat{c}_1^\dag-g_{ o}(\hat{b}_2+\hat{b}_2^\dag e^{2i\omega_mt})\nonumber\\
  &+\sqrt{2\kappa}\hat{c}_2^{\rm in}(t)-\frac{\lambda}{2\sqrt{2\kappa\eta_f}}(i\hat W_1(t)-\hat W_2(t)),
 \end{align}
 \label{les}
 \end{subequations}
where we have assumed $g_o=g_{oj}$ and $\omega_m=\omega_{mj}$ for simplicity, the operators $\hat{c}_j^{\rm in}$ represent vacuum noises entering the cavities and $\hat{b}_j^{\rm in} $ are thermal fluctuations of mechanical environments. We have the nonzero correlations $\langle \hat{b}_j^{\rm in}(t) \hat{b}_{j'}^{\rm in\dag}(t')\rangle=(\bar{n}_{th} +1)\delta_{jj'}\delta (t-t')$ and $\langle \hat{b}_j^{\rm in\dag}(t) \hat{b}_{j'}^{\rm in}(t')\rangle=\bar{n}_{th} \delta_{jj'}\delta (t-t')$, where  $\bar{n}_{ th}\equiv(e^{\frac{\hbar\omega_{m}}{k_BT}}-1)^{-1}$ is the mean number of thermal excitations of the mechanical environment at temperature $T$ and $k_B$ the Boltzmann constant. Note that in the above derivation the anti-RWA terms in Eq.(\ref{effem2}) have also been included. Indicated from Eq.(\ref{les}), the feedback not only modifies the cavity dissipation rate $\kappa$ and the NPD coupling strength $g_{p}$, but also introduce the detection noises which are correlated to the cavity input noises to the cavity fields. As will be shown below, the combination of these effects will alter the properties of the the cavity modes and the mechanical oscillators.

According to Eq.(\ref{les}), the equations of motion for the quadratures $\psi\equiv(\hat X_{b_1}, \hat Y_{b_1},\hat X_{b_2},\hat Y_{b_2},\hat X_{c_1},\hat Y_{c_1},\hat X_{c_2},\hat Y_{c_2})^T$ are given by
\begin{align}
\frac{d}{dt}\psi=-M_d \psi+\psi_{\rm in}(t),
\label{leq}
\end{align}
where the matrix $M_d=\left(
  \begin{array}{cc}
  M_{11} & M_{12} \\
  M_{21} & M_{22}
  \end{array}
  \right)$, $M_{11}=\frac{\gamma_m}{2}I_{4}$, $M_{22}=\left(
  \begin{array}{cc}
  (\kappa+\frac{\lambda}{2})I_2 & -(2g_p+\frac{\lambda}{2})\sigma_z \\
  -(2g_p+\frac{\lambda}{2})\sigma_z &  (\kappa+\frac{\lambda}{2})I_2
  \end{array}
  \right)$, $M_{12}=\tilde{M}_{12}I_2$, $\tilde M_{12}=g_o\left(
  \begin{array}{cc}
  -2\sin^2(\omega_m t) & \sin(2\omega_mt) \\
  \sin(2\omega_mt) & -2\sin^2(2\omega_m t)
  \end{array}
  \right)$, $M_{21}=\tilde{M}_{21}I_2$, $\tilde M_{21}=g_o\left(
  \begin{array}{cc}
  2\cos^2(\omega_m t) & -\sin(2\omega_mt) \\
  \sin(2\omega_mt) & 2\cos^2(2\omega_m t),
  \end{array}
  \right)$,
and $\psi_{\rm in}^T(t)=\big(\sqrt{\gamma_m}\hat X_{b_1}^{in}, \sqrt{\gamma_m}\hat Y_{b_1}^{in}, \sqrt{\gamma_m}\hat X_{b_2}^{in}, \sqrt{\gamma_m}\hat Y_{b_2}^{in}, \sqrt{2\kappa}\hat X_{c_1}^{in}+\frac{\hat W_2}{\sqrt{2\kappa\eta_f}}, \sqrt{2\kappa}\hat Y_{c_1}^{in}+\frac{\hat W_1}{\sqrt{2\kappa\eta_f}}, \sqrt{2\kappa}\hat X_{c_2}^{in}-\frac{\hat W_2}{\sqrt{2\kappa\eta_f}},\sqrt{2\kappa}\hat Y_{c_2}^{in}+\frac{\hat W_1}{\sqrt{2\kappa\eta_f}}\big)$, with $I_{j}$ the $j\times j$ identity matrix and $\sigma_z$ the $z$-component Pauli matrix.

Governed by Eq.(\ref{leq}), when the system  initially  starts from Gaussian states, it remains Gaussian during its evolution whose properties are determined by the second-order $8\times 8$ correlation matrix (CM) defined as $\sigma_{\rm om}^{ii'}=\langle \psi_i\psi_{i'}+ \psi_{i'}\psi_i\rangle/2$.
With Eqs.(\ref{nc1}), (\ref{nc2}) and (\ref{leq}), the CM of $\sigma_{\rm om}$ can be found to satisfy the following equation
  \begin{align}
 \frac{d}{dt} \sigma_{\rm om}=- M_d  \sigma_{\rm om}- \sigma_{\rm om} M_d^T+N_f,
 \label{ecm1}
  \end{align}
where the noise correlation matrix $ N_f=\left(
  \begin{array}{cc}
  N_{f_{11}} & 0 \\
  0 & N_{f_{22}}
  \end{array}
  \right)$, $N_{f_{11}}=\frac{\gamma_m}{2}(2\bar n_{th}+1)I_4$, and
\begin{align}
  N_{f_{22}}=\left(
  \begin{array}{cc}
  (\kappa+\frac{\lambda_f}{2}+\frac{\lambda_f^2}{8\kappa\eta_f})I_2 & -(\frac{\lambda_f}{2}+\frac{\lambda_f^2}{8\kappa\eta_f})\sigma_z \\
  -(\frac{\lambda_f}{2}+\frac{\lambda_f^2}{8\kappa\eta_f})\sigma_z &  (\kappa+\frac{\lambda_f}{2}+\frac{\lambda_f^2}{8\kappa\eta_f})I_2
  \end{array}
  \right).
  \end{align}
The above equation (\ref{ecm1}), which includes the time-dependent exponential terms, will be solved numerically. In the following discussion in Sec. \ref{sec4}, we also consider the case of RWA to obtain some analytical results. Under the RWA, the condition that all eigenvalues of the matrix $M_d$ are positive ensures the stability of Eq.(\ref{ecm1}). Simply, for the detection efficiency $\eta_f=1$ and negligible mechanical damping $\gamma_m=0$, the stability condition can be found to be
\begin{subequations}
\begin{align}
\kappa&>2g_p,\\
-(2g_p+\kappa)&<\lambda_f<(2g_p+\kappa).
\end{align}
\end{subequations}
In addition, the CM $\sigma_{\rm om}$ in the steady-state regime satisfies
\begin{align}
M_d  \sigma_{\rm om}+\sigma_{\rm om} M_d^T=N_f.
\label{scm}
  \end{align}
Thus, with the CM $\sigma_{\rm om}$ in the above Eq.(\ref{scm}), the $4\times 4$ CM $\sigma_{\rm m}$ of the subsystem of the two mechanical oscillators can be obtained. By using the CM $\sigma_{\rm m}$, one can discuss the properties of the entanglement, steering, and Bell nonlocality of the two optomechanical oscillators in the presence of quantum feedback.

\section{Measures for entanglement, steering, and Bell nonlocality}
Before discussion, we review in brief the measures for entanglement, steering, and Bell nonlocality of two mode Gaussian states. When the CM $\sigma_{m}$ is recast into the form $\sigma_{\rm m}=
  \left(
  \begin{array}{ccc}
  \sigma_{m}^{11} & \sigma_{m}^{12}\\
  \sigma_{m}^{12T} & \sigma_{m}^{22}
  \end{array}
 \right)$, where $\sigma_{m}^{ij}$ are $2\times 2$ matrices,
the entanglement can be quantified by the logarithmic negativity \cite{plen}
  \begin{align}
En=\max\big[0,-\ln(2\zeta_{\rm  en})\big].
  \end{align}
The parameter $\zeta_{\rm en}=\sqrt{s-\sqrt{s^2-4 \det \sigma_m}}$,
  where $s=\det \sigma_{m}^{11}+\det \sigma_{m}^{22}-2\det \sigma_{m}^{12}$. Hence, the parameter $\zeta_{\rm en}<\frac{1}{2}$ means the existence of entanglement.

For the subsystem of the two mechanical oscillators with the CM $\sigma_{m}$, the steering from the oscillator $\hat b_1$ ($\hat b_2$) to the oscillator $\hat b_2$ ($\hat b_1$) can be  measured by \cite{kog}
\begin{subequations}
 \begin{align}
  St_{1\rightarrow2}=\max\big[0,\frac{1}{2}\ln\frac{\det \sigma_{m}^{11}}{4\det \sigma_{m}}\big],\\
 St_{2\rightarrow1}=\max\big[0,\frac{1}{2}\ln\frac{\det \sigma_{m}^{22}}{4\det \sigma_{m}}\big].
  \end{align}
  \end{subequations}
When considering the symmetric situation, we set $St\equiv St_{1\rightarrow2}=St_{2\rightarrow1}$ and $\chi_{\rm st}\equiv\chi_{\rm st}^j=\frac{\det \sigma_{m}^{jj}}{4\det \sigma_{m}}$. Thus, $\chi_{st}>1$ shows the presence of the steering in two directions.

Based on displaced parity measurement operation, Banaszek and W\'{o}dkiewicz showed that Bell nonlocality of continuous-variable systems can be tested in phase space \cite{bana} by utilizing Wigner function of system which is the expectation value of the displaced parity operator in the state $\rho_b$, i.e., $W(\beta)=\frac{2}{\pi}\rm Tr[\hat \Pi(\beta)\hat \rho_b]$. Here, the operator
\begin{align}
\hat \Pi(\beta)&=\hat D_b(\beta)\sum_{n=0}^\infty\big(|2n\rangle\langle 2n |-|2n+1\rangle\langle 2n+1 |\big)\hat D_b^\dag(\beta),\nonumber\\
&=\hat D_b(\beta)(-1)^{\hat n}\hat D_b^\dag(\beta),
\end{align}
with the displacement operator $\hat D_b(\beta)=\exp (\beta \hat b^\dag-\beta^*\hat b)$ for a bosonic operators $\hat b$ and $\hat b^\dag$, and $\hat n=\hat b^\dag\hat b$. For a two-mode bosonic field described by the annihilation operators $\hat b_1$ and $\hat b_2$, by constructing the joint parity measurement
\begin{align}
\hat \Pi(\beta_1,\beta_2)=\hat D_{b_1}(\beta_1)\hat D_{b_2}(\beta_2)(-1)^{\hat n_1+\hat n_2}\hat D_{b_1}^\dag(\beta_1)\hat D_{b_2}^\dag,
\end{align}
where $\hat n_j=\hat b_j^\dag\hat b_j$,  and utilizing the two-mode field Wigner function $W (\beta_1, \beta_2)=\frac{4}{\pi^2}\rm Tr\big[ \hat \Pi(\beta_1,\beta_2)\hat \rho_{b_1b_2}\big]$, one can derive that the Bell-Clauser-Horne-Shimony-Hold (CHSH) inequality imposed by local hidden theory can be expressed as \cite{bana}
\begin{align}
B(\beta_1,\beta_2)=&\frac{4}{\pi^2}\big[W(\beta_1,\beta_2)+W(\beta'_1,\beta_2)+W(\beta_1,\beta'_2)\nonumber\\
&-W(\beta'_1,\beta'_2)\big].
\end{align}
Local hidden theory imposes $|B|\leq 2$ and the maximal violation allowed by quantum mechanics is $|B|_{\rm max}=2\sqrt{2}$.
For the present system of two mechanical oscillators, the Wigner function
\begin{align}
  W(\mu)=\frac{\exp{(-\mu\sigma_{\rm m}^{-1}\mu^T)}}{\pi^2\sqrt{\det\sigma_{\rm m}}},
\end{align}
where $\mu=\{\beta_{1x}, \beta_{1y},\beta_{2x}, \beta_{2y}\}$, with $\beta_{jx}=\text {Re}[\beta_j]$ and $\beta_{jy}=\text {Im}[\beta_j]$. Thus, when obtaining the CM $\sigma_{\rm m}$, we can calculate the maximal value $|B|_{\rm max}$ via optimizing $B$ over the full range of the phase-space variables $\{\beta_{1x}, \beta_{1y},\beta_{2x}, \beta_{2y},\beta'_{1x}, \beta'_{1y},\beta'_{2x}, \beta'_{2y}\}$.

\section{Results and discussion}
\label{sec4}
\begin{figure}[t]
\centerline{\scalebox{0.38}{\includegraphics{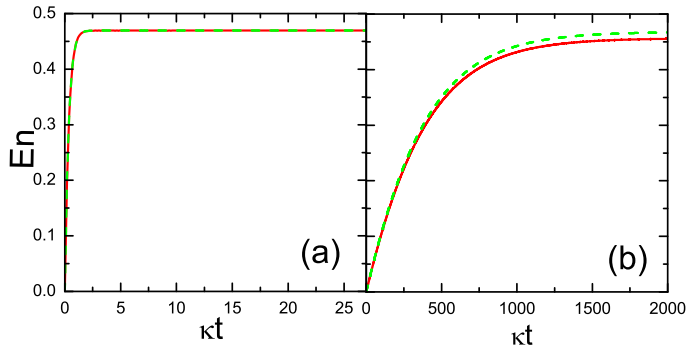}}}
 \caption{The time dependence of the intracavity-field entanglement (a) and the mechanical entanglement (b) in the absence of the feedback ($\lambda_f=0$) for the mechanical frequency $\omega_m=10\kappa$, the NPD coupling strength $g_p=0.3\kappa$, the optomechanical coupling strength $g_o=0.05\kappa$, the mechanical damping rate $\gamma_m=10^{-5}\kappa$, and the mean number of thermal phonons $\bar {n}_{\rm th}=0$. The dashed and solid lines correspond to the cases with and without the RWA, respectively.}
 \label{fig2}
\end{figure}
\begin{figure}[t]
\centerline{\scalebox{0.25}{\includegraphics{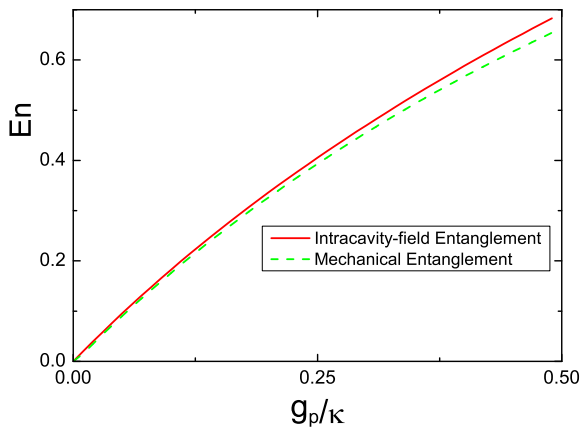}}}
 \caption{The dependence of the steady-state intracavity-field entanglement and the mechanical entanglement on the NPD coupling $g_p$ without the feedback. The other parameters are the same as in Fig.\ref{fig2}}.
 \label{fig3}
\end{figure}
\subsection{Without the feedback}
In this section, we study in detail the properties of the entanglement, steering and Bell nonlocality of the two mechanical oscillators. We firstly consider the situation in the absence of the feedback, i.e., $\lambda_f=0$. In Fig.2 (a) and (b), the time evolution and long-time behavior for the intracavity-field entanglement and mechanical entanglement are plotted, respectively. It is explicitly shown that the degree of the mechanical entanglement in the steady-state regime is approximately equal to that of the intracavity field. This is because that the cavity-field entanglement, which is built up via the NPD, is transferred to the mechanical oscillators with the help of the local optomechanical linear mixing [the first term in Eq.(\ref{linh})].
However, due to the cavity dissipation, which is also necessary to achieve the steady states for negligible mechanical damping, there thus exists a little difference between the two steady-state entanglement degrees, as shown in Fig.2.
Likewise, for the same reason it can be seen that the long-time intracavity-field entanglement is less affected by the anti-RWA terms than the mechanical entanglement for the optomechanical coupling $g_o\ll \kappa$. The existence of the anti-RWA terms deceases the mechanical entanglement in the long-time regime. In addition, for $\kappa\gg g_{o}$ and $\gamma_m\approx0$, the cavity-field achieves the steady-state entanglement much faster than the couple of the mechanical oscillators, which thus allows us to adiabatically eliminate the cavity modes to obtain approximate analytical results presented in the following. In Fig.3, the long-time intracavity-field and mechanical entanglement are plotted. From it one can see that the maximal long-time entanglement $En\approx 0.68$, which occurs at the coupling $g_p\approx\frac{\kappa}{2}$, under the constraint of the stability condition $\kappa>2g_p$ for the mechanical damping rate $\gamma_m\approx0$. This is because that for an intracavity NPD, the maximal steady-state entanglement $En=\ln 2$, which occurs on the threshold ($g_p=\frac{\kappa}{2}$), i.e., the so-called 3~dB limit. It should be noted that for the NPD at this limit, the steerable correlations can not be achieved for the balanced dissipation rates of the two cavity modes $\hat c_j$ \cite{tan}. Therefore, merely with the NPD inside the optomechanical cavity, the mechanical steering and Bell nonlocality achieved with stronger correlations are unattainable.
\begin{figure*}[t]
\centerline{\scalebox{0.28}{\includegraphics{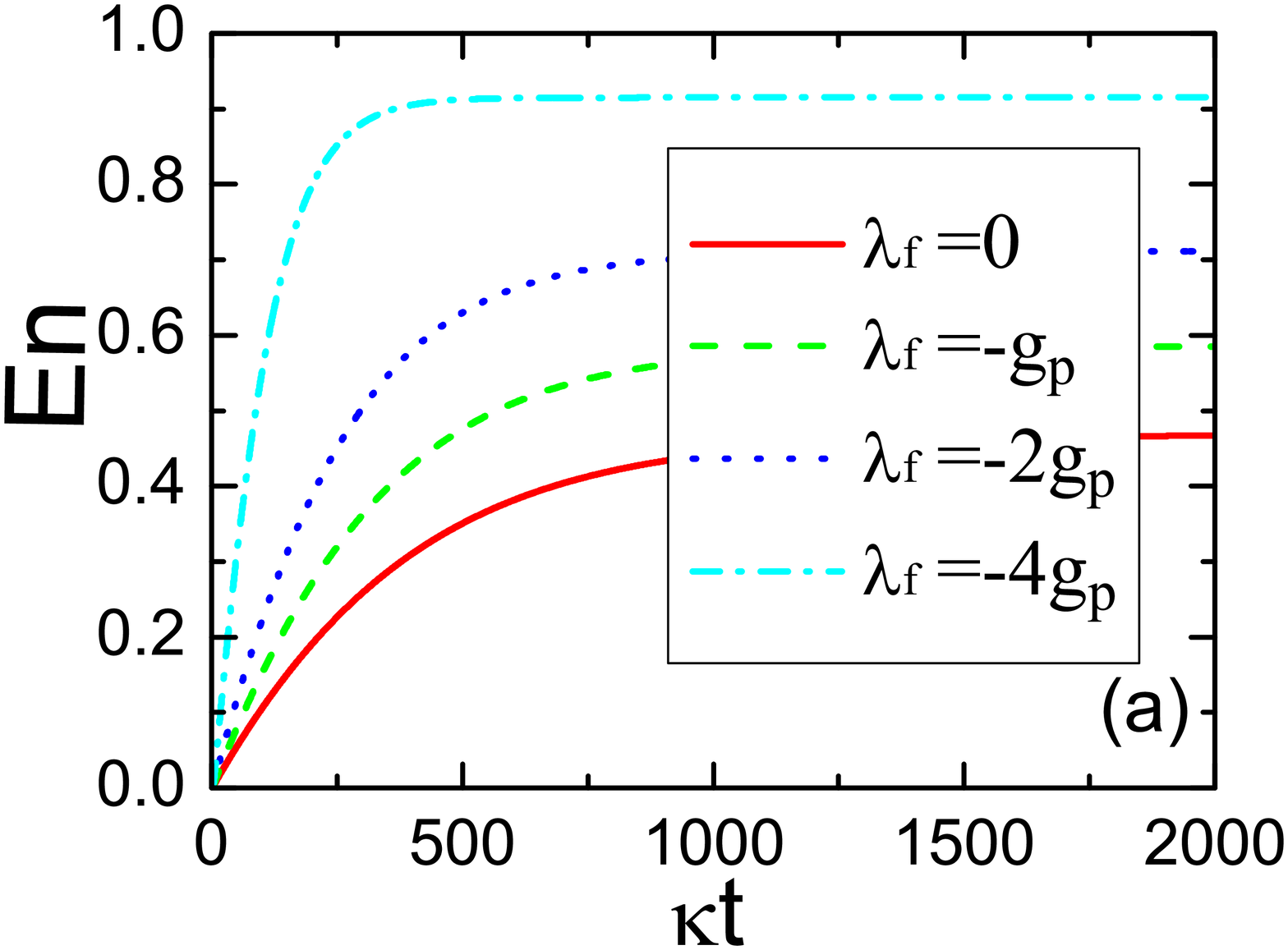}~\includegraphics{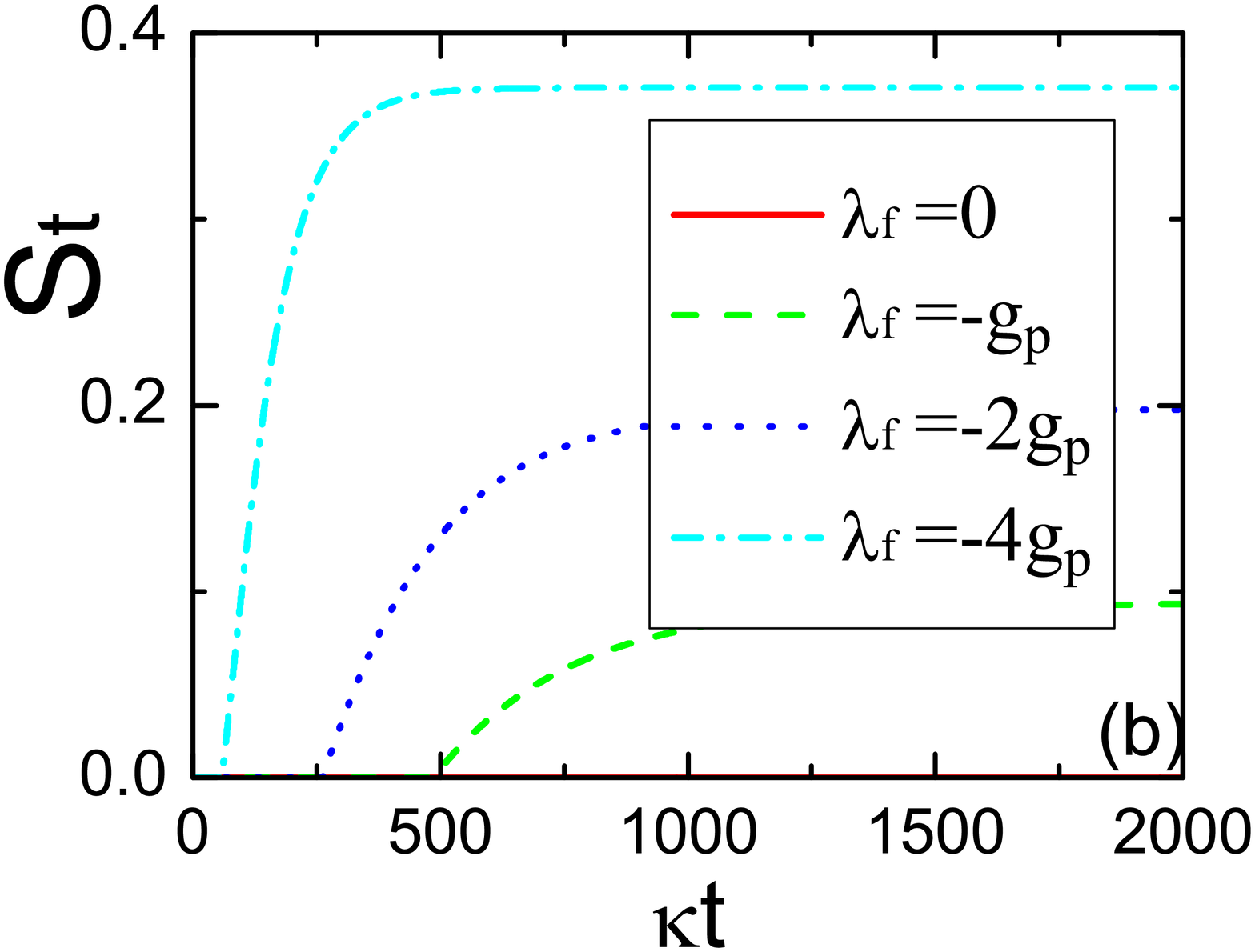}}}
\centerline{\scalebox{0.28}{\includegraphics{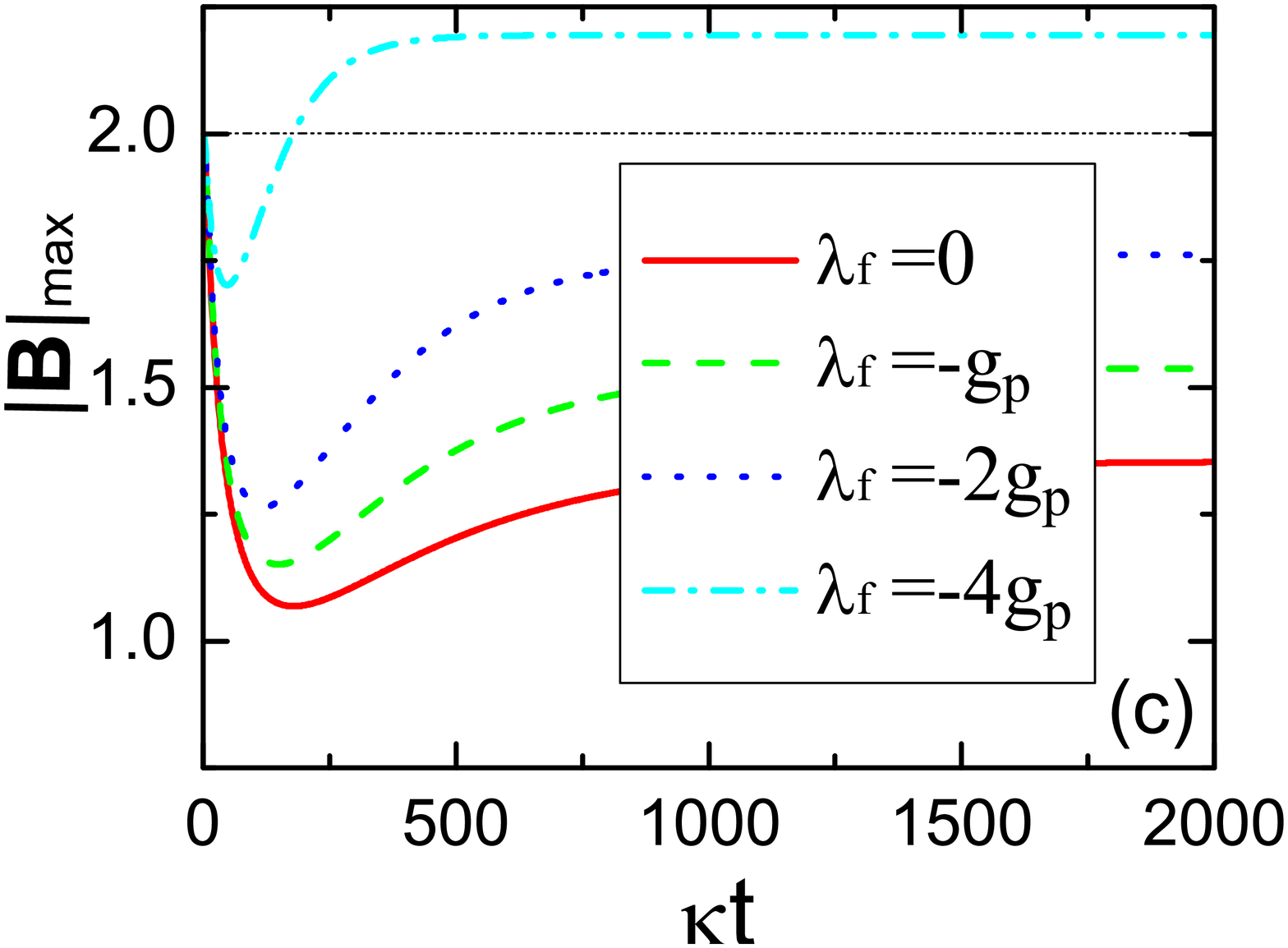}~\includegraphics{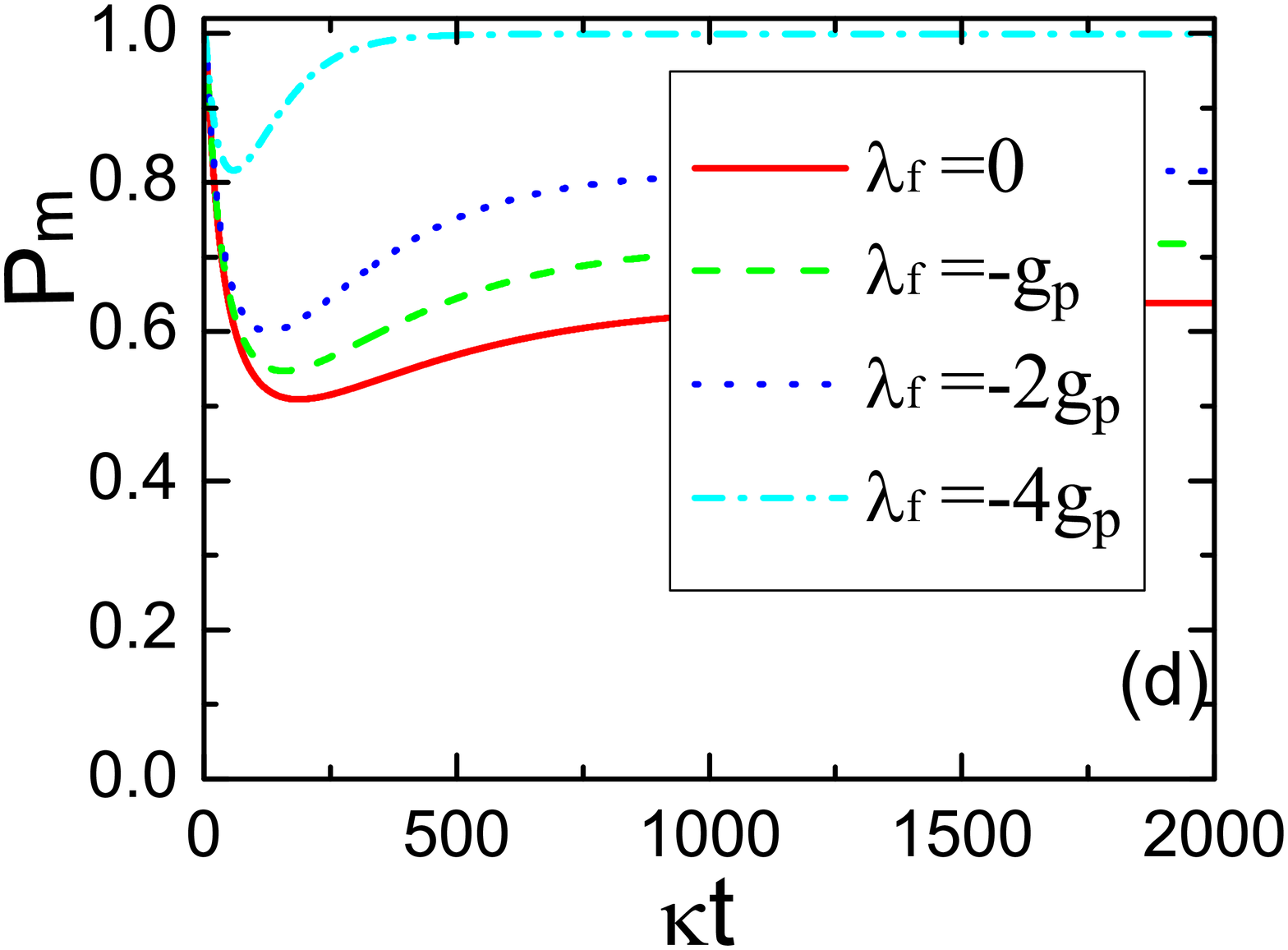}}}
 \caption{The time dependenc of the mechanical entanglement $En$, steering $St$, Bell nonlocality $|B|_{\rm max}$, and purity $P_m$ of the mechanical states for different values of the feedback strength $\lambda_f$. The other parameters are the same as in Fig.2.}
 \label{fig4}
\end{figure*}
\subsection{With the feedback}
We next consider the generation of the mechanical steering and Bell nonlocality by exploiting quantum feedback. In Fig.4, the time development of the entanglement $En$, steering $St$, Bell nonlocality $|B|_{\rm max}$ and purity $P_m=Tr(\hat \rho_m^2)=\frac{1}{4\sqrt{\det \sigma_m}}$ of the mechanical states $\hat \rho_m$ is plotted for different feedback strengthes $\lambda_f$, with $\eta_f=1$. It is shown that the feedback can effectively enhance the mechanical entanglement and realize the mechanical steering and Bell nonlocality. Moreover, the purity of the mechanical states can also be effectively increased by the feedback. In addition, one can also see that the introduction of the feedback leads to shorter time scale on which the system reaches the steady states. In Fig.5, we plot the dependence of the mechanical entanglement, steering and Bell nonlocality, and purity in the long-time regime  on the feedback strength $\lambda_f$. We can see from it that as the feedback strength increases, the mechanical entanglement, steering and purity at first increases and then decreases. The enhancement merely occurs in the region where the feedback $\lambda_f<0$. The peaks of the entanglement, steering and purity locate around the feedback strength $\lambda_f\approx -4g_p$ and they increase with the increasing of the NPD coupling strength $g_p$. It is also shown that Bell nonlocality is presented in the vicinity of $\lambda_f\approx-4g_p$, and for such a feedback strength the purity $P_m\approx1$, meaning pure entangled steerable mechanical states. In fig.6, the dependence of the entanglement, steering, Bell nonlocality and purity on $g_p$ for the feedback $\lambda_f=-4g_p$. We can see that the entanglement and steering increase as $g_p$ increases. The purity $P_m\approx1$ and the Bell nonlocality exists in the whole range of $g_p\in (0, \frac{\kappa}{2})$. The feedback leads to the maximal degrees of the entanglement and steering when the NPD operates near the threshold, i.e., $g_p\approx\frac{\kappa}{2}$.

From the above discussion, we can therefore conclude that: (i) in the absence of the feedback, only mechanical entanglement in the long-time regime can be obtained; (ii) with the feedback the mechanical entanglement can be enhanced considerably and the mechanical steering and Bell nonlocality can be achieved; (iii) the entanglement, steering and Bell nonlocality peak at the feedback strength $\lambda_f=-4g_p$ for which the purity of the entangled steerable mechanical states $P_m\approx1$, for negligible mechanical damping; (iv) the peaks increase with the increasing of the NPD coupling $g_p$ and become maximal when the NPD operates near the threshold, i.e., $g_p\approx \frac{k}{2}$.

\subsection{Approximate analytical results}
To understand the above results, we consider the adiabatical elimination of the two cavity modes from
the dynamics of the mechanics under the condition that $\omega_m\gg\kappa\gg g_o$
which can also allows us to perform the RWA. This can be done by letting $\frac{d}{dt}\hat c_j=0$ in Eq.( \ref{les}),
discarding the time-dependent exponential terms,  and
substituting the expressions of $\hat{c}_1$ and $\hat{c}_2$ into the motion equations of mechanical oscillators $\hat{b}_1$ and $\hat{b}_2$.  The resulting equations are given by
  \begin{subequations}
 \begin{align}
 \frac{d}{dt}\hat{b}_1=&-\Big(\frac{\gamma_m}{2}+\gamma_{\rm eff}\Big)\hat{b}_1
 -G_{\rm eff}\hat{b}_2^{\dag}+\sqrt{2\gamma_{\rm eff}}\hat {\tilde b}_1^{\rm in}+\sqrt{\gamma_m}\hat{b}_1^{\rm in},\\
 \frac{d}{dt}\hat{b}_2=&-\Big(\frac{\gamma_m}{2}+\gamma_{\rm eff}\Big)\hat{b}_2
 -G_{\rm eff}\hat{b}_1^{\dag}+\sqrt{2\gamma_{\rm eff}}\hat {\tilde b}_2^{\rm in}
 +\sqrt{\gamma_m}\hat{b}_2^{\rm in},
 \end{align}
 \label{effem}
\end{subequations}
where $G_{\rm eff}=\frac{g_o^2(2g_p+\frac{\lambda}{2})}{(\kappa-2g_p)(\kappa+2g_p+\lambda)}$, $\gamma_{\rm eff}=\frac{g_o^2(\kappa+\frac{\lambda}{2})}{(\kappa-2g_p)(\kappa+2g_p+\lambda)}$, and the new noise operators $\hat{\tilde{b}}_j^{\rm in}$
\begin{subequations}
 \begin{align}
 \hat{\tilde{b}}_1^{\rm in}=\cosh r\hat{c}_1^{\rm in}+\sinh r \hat{c}_2^{ \rm in\dag},\\
\hat{ \tilde{b}}_2^{\rm in}=\sinh r\hat{c}_1^{\rm in\dag}+\cosh r \hat{c}_2^{\rm in},
 \end{align}
\end{subequations}
with the squeezing parameter
\begin{align}
r=\tanh^{-1}\Big[\frac{4\kappa(2g_p+\frac{\lambda_f}{2})-\lambda_f(\kappa-2g_p)}{4\kappa(\kappa+\frac{\lambda_f}{2})+\lambda_f(\kappa-2g_p)}\Big].
\label{sf}
\end{align}
It is obvious that the noise operators satisfy the following nonvanishing correlations
\begin{subequations}
 \begin{align}
 &\langle \hat{\tilde{b}}_j^{\rm in\dag}(t)\hat{\tilde{b}}_j^{\rm in}(t')\rangle=\sinh^2r \delta(t-t'),\\
 &\langle \hat{\tilde{b}}_j^{\rm in}(t)\hat{\tilde{b}}_j^{\rm in\dag}(t')\rangle=\cosh^2r\delta(t-t'),\\
 &\langle \hat{\tilde{b}}_1^{\rm in}(t)\hat{\tilde{b}}_2^{\rm in}(t')\rangle=\sinh r \cosh r \delta(t-t').
 \end{align}
 \label{nc3}
\end{subequations}
It can be seen from Eqs.(\ref{effem}) and (\ref{nc3}) that the combination of the optomechanical coupling (related to $g_o$) and the feedback can effectively enhance the mechanical damping, with the effective damping rate $\gamma_{\rm eff}$. That is why the mechanical subsystem approaches faster the steady states in the presence of the feedback, as depicted in Fig.\ref{fig4}. The optomechanical coupling to the cavity modes also gives rise to the effective NPD between the two mechanical oscillators, with the strength $G_{\rm eff}$ which is in turn modified by the feedback. Furthermore, the feedback introduces a broadband two-mode squeezed vacuum bath (described by Eq.(\ref{nc3})) which is coupled to the two mechanical oscillators. Therefore, Eq.(\ref{effem}) effectively describes a mechanical NPD of which the two mechanical modes are immersed in a two-mode squeezed vacuum environment for the case that the mechanical damping rate $\gamma_m=0$.

It can be found from Eq.(\ref{effem}) that the entanglement and steering parameters
\begin{align}
\zeta_{\rm en}&=\frac{(2\kappa+\lambda_f)^2}{8\kappa(2g_p+\kappa+\lambda_f)}
+\frac{\gamma_m(2\bar{n}_{\rm th}+1)(2g_p+\kappa+\lambda_f)}{4g_o^2},
\end{align}
\begin{widetext}
\begin{align}
\chi_{\rm st}^{\frac{1}{2}}&=\frac{\Big[(4\kappa+\lambda_f)\big[\kappa(2\kappa+\lambda_f)-2 g_p\lambda _f \big]\Big]g_o^4+2\kappa\gamma_m(2\bar {n}_{\rm th}+1)(2\kappa+\lambda_f)(\kappa-2g_p)(2g_p+\kappa+\lambda_f)g_o^2}{\big[2\kappa g_o^2+\gamma_m(2\bar {n}_{\rm th}+1)(\kappa-2g_o)^2\big]\big[2\kappa\gamma_m(2\bar {n}_{\rm th}+1)(2g_p+\kappa+\lambda_f)^2+g_o^2(2\kappa+\lambda_f)^2\big]}.
\label{26}
\end{align}
\end{widetext}
When the mechanical damping $\gamma_m=0$, Eq.(\ref{26}b) reduces to
\begin{align}
\chi_{\rm st}&=1-\frac{(\kappa+2g_p)\lambda_f^2+8 \kappa g_p\lambda_f}{2\kappa(2\kappa+\lambda_f)^2},
\end{align}
which is independent of the optomechanical coupling $g_o$.
Hence, we can see that when the mechanical damping is negligible, the steady-state entanglement and steering is independent of the coupling $g_o$ under the condition that $\omega_m\gg \{g_o, \kappa, \gamma_m\}$ such that the anti-RWA terms can be discarded. While with the finite mechanical damping rate, the increasing of the coupling $g_o$ decreases the entanglement and steering in the long-time regime.  For $\gamma_m\approx0$ and $\lambda_f=0$, the entanglement and steering parameters in the above become into
\begin{subequations}
\begin{align}
\zeta_{\rm en}&=\frac{\kappa}{2(2g_p+\kappa)},\\
\chi_{\rm st}&=1.
\end{align}
\label{27}
\end{subequations}
It is shown from Eq.(\ref{27}a) that the mechanical entanglement, achieved via the effective mechanical NPD, increases as the increasing of the coupling $g_p$  and decreasing of the dissipation rate $\kappa$. Near the threshold $g_p\approx \frac{\kappa}{2}$, we have $\zeta_{\rm en}\approx \frac{1}{4}$ and thus the maximal mechanical entanglement in the steady-state regime $En\approx\ln 2$. This is the so-called $3\rm~dB$ limit for the maximally attainable entanglement of a NPD in the steady state regime which is transferred to the mechanical oscillators from the cavity fields in the present scheme. As for the steering, Eq.(\ref{27}b) shows that it is unachievable in the absence of the feedback,
different from the behavior of the entanglement.

\begin{figure}[t]
\centerline{\scalebox{0.50}{\includegraphics{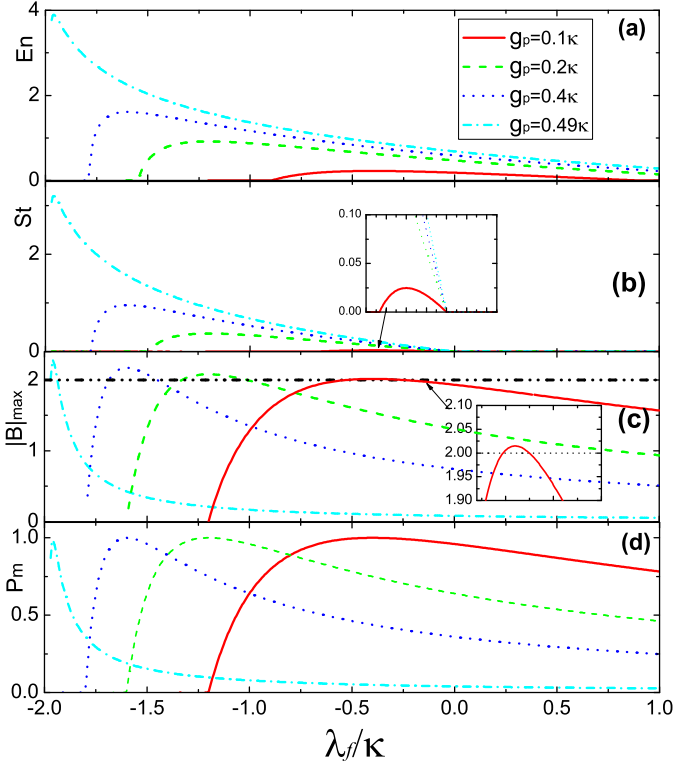}}}
 \caption{The dependence of the mechanical entanglement $En$, steering $St$, Bell nonlocality $|B|_{\rm max}$, and purity $P_m$ of the mechanical states in the long-time regime on the feedback strength $\lambda_f$ for different values of NPD coupling $g_p$. The other parameters are the same as in Fig.2.}
 \label{fig5}
\end{figure}

\begin{figure}[t]
\centerline{\scalebox{0.25}{\includegraphics{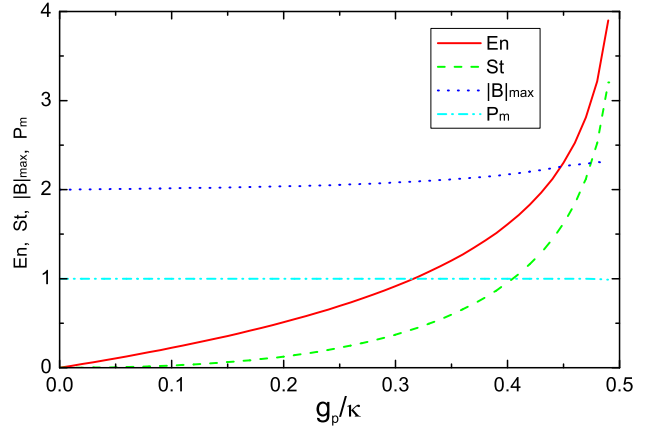}}}
 \caption{The dependence of the mechanical entanglement $En$, steering $St$, Bell nonlocality $|B|_{\rm max}$, and purity $P_m$ of the mechanical states in the long-time regime on the NPD coupling for the feedback strength $\lambda_f=-4g_p$, and the other parameters are the same as in Fig.\ref{fig2}.}
 \label{fig6}
\end{figure}
\begin{figure}[t]
\centerline{\scalebox{0.30}{\includegraphics{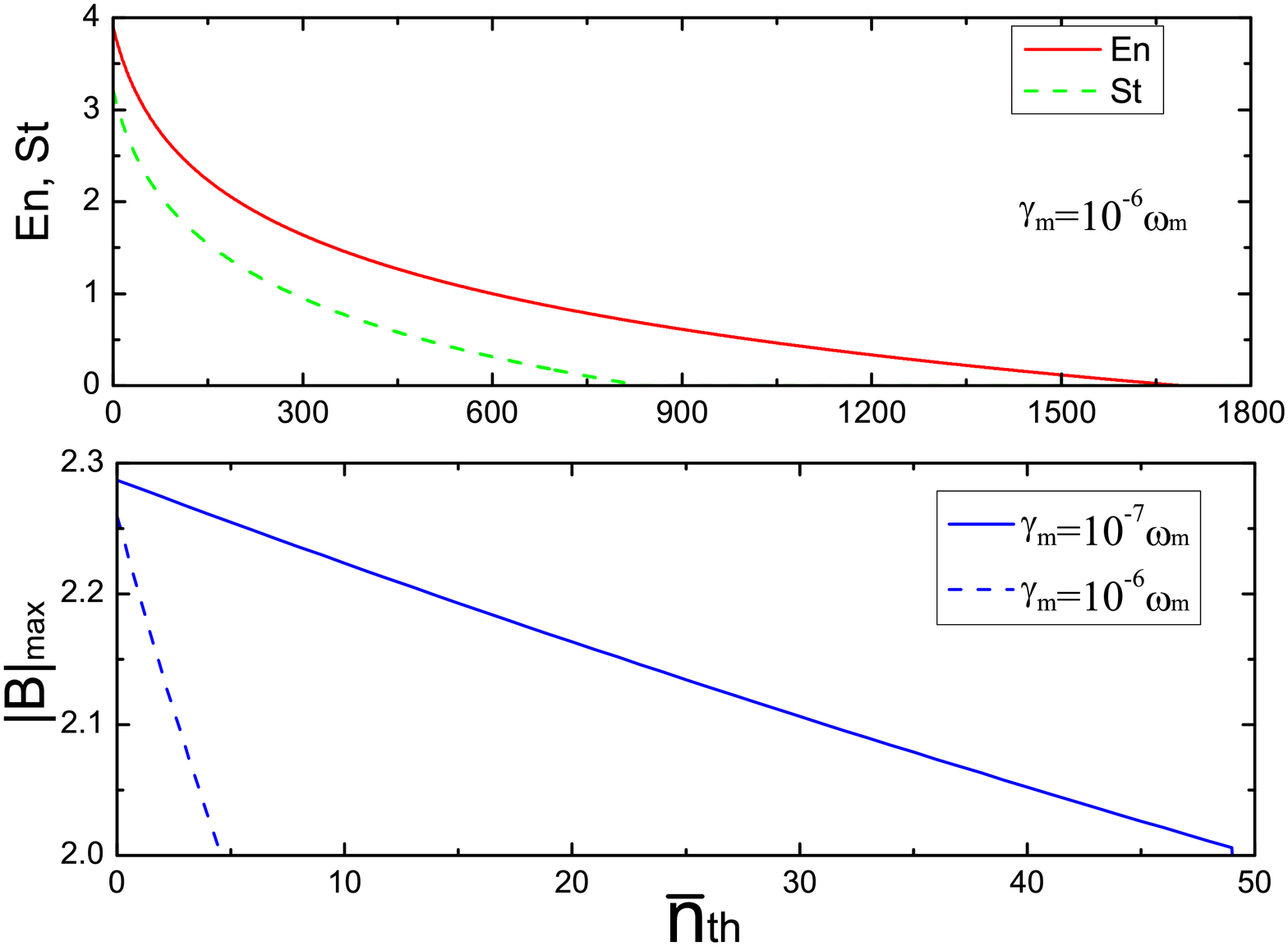}}}
\caption{The effects of thermal fluctuations on the mechanical entanglement, steering, and Bell nonlocality, for the NPD coupling $g_p=0.49\kappa$, and the other parameters are the same as in Fig.6.}
 \label{fig8}
\end{figure}

When the feedback is present, indicated from Eqs.(\ref{26}a) and (\ref{26}b), for $\gamma_m=0$ the derivatives $\frac{d\zeta_{\rm en}}{d\lambda_f}=\frac{(\lambda_f+4g_p)(\lambda_f+2\kappa)}{8\kappa(2g_p+\kappa+\lambda_f)}$ and $\frac{d\chi_{\rm st}}{d\lambda_f}=-\frac{2\kappa(\lambda_f+4g_p)}{(\lambda_f+2\kappa)^3}$. Thus, as depicted in Fig.\ref{fig5}, the mechanical entanglement and steering decreases as the feedback strength increases when $\lambda_f>0$. While $\lambda_f<0$ and $-4g_p<\lambda_f<-2\kappa$, the mechanical entanglement and steering increases at first as the feedback strength $(-\lambda_f)$ increases and then deceases as it continues to increase. Further, when the strength $\lambda_f\approx-4g_p$, the derivations $\frac{d\zeta_{\rm en}}{d\lambda_f}=0$ and $\frac{d\chi_{\rm st}}{d\lambda_f}=0$, which means that the steady-state entanglement and steering become maximum, with respect to the feedback strength $\lambda_f$. In addition, for this value of $\lambda_f$, the entanglement and steering parameters
\begin{subequations}
\begin{align}
\zeta_{\rm en}&\approx\frac{\kappa-2g_p}{2\kappa},\\
\chi_{\rm en}&\approx 1+\frac{2g_p^2}{\kappa(\kappa-2g_p)}.
\end{align}
\end{subequations}
Near the threshold $g_p\approx\frac{\kappa}{2}$, we have $\zeta_{\rm en}\approx 0$ and $\chi_{\rm en}\approx \infty$, and thus
$En\approx\infty~~\text{and}~~St\approx \infty$, corresponding to original (perfect) EPR entangled state \cite{EPR} which can never be obtained realistically because they corresponds to the situation of infinite energy (i.e., $\langle \hat b_j^\dag \hat b_j\rangle\approx \infty$). Thus, as shown in Fig.\ref{fig6}, the feedback-driven entanglement and the steering become maximal when the feedback strength $\lambda_f\approx-4g_p$ and the NPD coupling strength $g_p\approx \frac{\kappa}{2}$. This is because when the feedback strength $\lambda_f\approx-4g_p$ in Eq.(\ref{effem}) the effective mechanical NPD coupling $G_{\rm eff}\approx0$ and the mechanical oscillators are then directly coupled to a broadband squeezed vacuum, i.e.,
\begin{subequations}
 \begin{align}
 \frac{d}{dt}\hat{b}_1=&-\gamma_{\rm eff}\hat{b}_1+\sqrt{2\gamma_{\rm eff}}\hat {\tilde b}_1^{\rm in},\\
 \frac{d}{dt}\hat{b}_2=&-\gamma_{\rm eff}\hat{b}_2
 +\sqrt{2\gamma_{\rm eff}}\hat {\tilde b}_2^{\rm in}.
 \end{align}
 \label{effem1}
\end{subequations}
When performing a two-mode squeezing transformation $\hat S(r)=\exp[-r(\hat b_1\hat b_2-\hat b_1^\dag\hat b_2^\dag)]$ on the above equations, with the transformation relations $\hat S^\dag\hat b_1\hat S=\cosh r \hat b_1+\sinh r \hat b_2^\dag$ and $\hat S^\dag\hat b_2\hat S=\cosh r \hat b_2+\sinh r \hat b_1^\dag$, one has the resulting equation
\begin{subequations}
 \begin{align}
 \frac{d}{dt}\hat{b}_1=&-\gamma_{\rm eff}\hat{b}_1+\sqrt{2\gamma_{\rm eff}}\hat {c}_1^{\rm in},\\
 \frac{d}{dt}\hat{b}_2=&-\gamma_{\rm eff}\hat{b}_2
 +\sqrt{2\gamma_{\rm eff}}\hat {c}_2^{\rm in}.
 \end{align}
 \label{effem1}
\end{subequations}
In the transformed picture, the modes $\hat b_1$ and $\hat b_2$ evolve asymptotically into vacuum $|0_{ b_1}0_{ b_2}\rangle_{ss}$, where the subscript ``ss" denotes steady states and therefore in the original picture, the mechanical oscillators are prepared in a two-mode squeezed vacuum, i.e.,
\begin{align}
|\psi_{b_1b_2}\rangle_{ss}=\hat S(r)|0_{b_1}0_{b_2}\rangle.
\end{align}
Hence, as shown in Figs.\ref{fig5} and \ref{fig6}, the purity of the mechanical states $P_m\approx 1$ when the feedback strength $\lambda_f\approx-4g_p$. For such a feedback strength, the squeezing parameter $r$ in Eq.(\ref{sf}) reduces to
 \begin{align}
 r=\tanh^{-1}\Big[\frac{g_p}{\kappa-g_p}\Big].
 \end{align}
As shown in Fig.\ref{fig6}, with the increasing of $g_p$, the entanglement and steering increases. Moreover, for such a value of the feedback strength and when NPD coupling strength $g_p\approx\frac{\kappa}{2}$, we have the squeezing factor in Eq.(\ref{sf}) $r\approx\infty$, which corresponds to the perfect EPR state, and therefore theoretically the mechanical entanglement and steering $En\approx\infty$ and $St\approx \infty$.

Finally, we consider the effect of thermal fluctuations on the mechanical entanglement, steering and Bell nonlocality, which is shown in Fig.\ref{fig8}. Since the steering is intermediate between the entanglement and Bell nonlocality,  we see that the maximal number $\bar n_{th}$ up to which the steering still exist is also intermediate between those for the entanglement and Bell nonlocality. The Bell nonlocality is much more vulnerable than the entanglement and steering to the thermal effect. As shown in Fig.\ref{fig8}, for the mechanical quality factor $Q_m=\omega_m/\gamma_m=10^6$, the maximal thermal phonon number $\bar n_{th}\approx 5$. If we accept the parameters close to those in Ref.\cite{weaver} that the cavity dissipation rates $\kappa/2\pi\approx100$ MHz, the mechanical frequency $\omega_m/2\pi\approx 1$ GHz and $\gamma_m/2\pi\approx 1$ kHz , the the corresponding temperature $T\approx 40$mK. For higher mechanical quality factor $Q_m=10^7$ of, e.g., vibrating membranes used in the experiments in Ref.\cite{har}, the corresponding maximal thermal phonon number $\bar n_{th}\approx 50$ and thus the temperature $T\approx 0.4$ K. Hence, precooling to the ground states and high quality of the mechanical oscillators are necessary for achieving Bell nonlocality.

\section{Conclusion}
To summarize,  here we propose a scheme for realizing macroscopic quantum steering and Bell nonlocality of two optomechanical oscillators in the steady-state via continuous quantum measurement and feedback. In the present cavity optomechanical system, two mechanical oscillators are dispersively coupled to two cavity fields which in turn interact to each other via an intracavity NPD. We consider that the two cavity output fields are subject to the Bell-like homodyne detection and the detection currents are fed back to drive the cavity fields and thus modify the dynamics of the whole system. We show without the feedback, the two mechanical oscillators can be prepared in steady entangled states via the NPD but the mechanical entangled states do not exhibit quantum steering and Bell nonlocality. When the feedback is present, it is found that with the help of the NPD interaction, the two mechanical oscillators can be driven by the feedback into an approximate two-mode squeezed vacuum steady state which displays strong quantum steerable correlations and Bell nonlocality. It is shown that the mechanical Bell nonlocality is much more sensitive to thermal decoherence than the mechanical steering. Therefore, to achieve the mechanical Bell nonlocality, high-quality mechanical resonators and precooling are necessary. Apart from the application in fundamental test of quantum mechanics, the macroscopic steering and Bell nonlocality may also be used in quantum informatics.

\section*{Acknowledgment}
This work is supported by the National Natural Science Foundation of China (No.11674120), and the Fundamental Research Funds for the Central Universities (No. CCNU18TS033).

\end{document}